\documentclass[conference]{IEEEtran}
\IEEEoverridecommandlockouts

\usepackage{cite}
\usepackage{amsmath,amssymb,amsfonts}
\usepackage{algorithmic}
\usepackage{graphicx}
\usepackage{textcomp}
\usepackage[table,xcdraw,dvipsnames]{xcolor}
\def\BibTeX{{\rm B\kern-.05em{\sc i\kern-.025em b}\kern-.08em
    T\kern-.1667em\lower.7ex\hbox{E}\kern-.125emX}}

\usepackage[normalem]{ulem} 
\useunder{\uline} {\ul} {} 
\usepackage{kotex} 
\usepackage{ulem} 
\usepackage{booktabs} 
\usepackage{multirow} 
\usepackage{subfigure} 
\usepackage{url} 
\usepackage{array} 
\usepackage{caption} 
\usepackage{fancyhdr}
\usepackage{xcolor}
\usepackage{enumitem}
\usepackage{bbm}
\usepackage{tabularx}
\usepackage[linesnumbered,ruled,vlined]{algorithm2e}

\newcolumntype{L} [1]{>{\raggedright\let\newline\\\arraybackslash\hspace{0pt} } m{#1} } 
\newcolumntype{C} [1]{>{\centering\let\newline\\\arraybackslash\hspace{0pt} } m{#1} } 
\newcolumntype{R} [1]{>{\raggedleft\let\newline\\\arraybackslash\hspace{0pt} } m{#1} } 

\newcolumntype{S}[1]{>{\raggedleft\arraybackslash}m{#1}}
\newcolumntype{P}[1]{>{\arraybackslash}m{#1}}
\newcolumntype{E}[1]{>{\color[rgb]{0.55,0.55,0.55}\scriptsize\arraybackslash}m{#1}}

\begin{document}

\title{Layer-aware TDNN: Speaker Recognition Using Multi-Layer Features from Pre-Trained Models
\thanks{This research was supported by the BK21 FOUR funded by the Ministry of Education of Korea and National Research Foundation of Korea. This research was also results of a study on the ``Leaders in INdustry-university Cooperation 3.0'' Project, supported by the Ministry of Education and National Research Foundation of Korea.}
}

\author{
    \IEEEauthorblockN{
        Jin Sob Kim,
        Hyun Joon Park,
        Wooseok Shin,
        Juan Yun,
        Sung Won Han
    }
    \IEEEauthorblockA{
        \textit{Department of Industrial and Management Engineering} \\
        \textit{Korea University}\\
        Seoul, Republic of Korea \\
        \{jinsob, winddori2002, wsshin95, yunjuan, swhan\}@korea.ac.kr
    }
}

\maketitle

\begin{abstract}
Recent advances in self-supervised learning (SSL) on Transformers have significantly improved speaker verification (SV) by providing domain-general speech representations.
However, existing approaches have underutilized the multi-layered nature of SSL encoders.
To address this limitation, we propose the layer-aware time-delay neural network (L-TDNN), which directly performs layer/frame-wise processing on the layer-wise hidden state outputs from pre-trained models, extracting fixed-size speaker vectors.
L-TDNN comprises a layer-aware convolutional network, a frame-adaptive layer aggregation, and attentive statistic pooling, explicitly modeling of the recognition and processing of previously overlooked layer dimension.
We evaluated L-TDNN across multiple speech SSL Transformers and diverse speech-speaker corpora against other approaches for leveraging pre-trained encoders.
L-TDNN consistently demonstrated robust verification performance, achieving the lowest error rates throughout the experiments.
Concurrently, it stood out in terms of model compactness and exhibited inference efficiency comparable to the existing systems.
These results highlight the advantages derived from the proposed layer-aware processing approach.
Future work includes exploring joint training with SSL frontends and the incorporation of score calibration to further enhance verification state-of-the-art performance.
\end{abstract}


\section{Introduction} \label{sec:introduction}
Speaker verification (SV) authenticates an identity by extracting speaker-specific features from speech.
The field has advanced significantly due to deep neural network (DNN) improvements and enlarged data resources.
One of the earliest, the $x$-vector \cite{snyder_2018_xvector} established a foundational DNN-based architecture upon handcrafted acoustic features (e.g., MFCCs).
It comprises a stack of time-delay neural network (TDNN) \cite{waibel_1989_tdnn} layers, temporal pooling, and utterance-level processing.
The pipeline has been refined with deeper TDNNs \cite{snyder_2019_etdnn, yu_2020_dtdnn, zhang_2020_aret, desplanques_2020_ecapatdnn}, attention-based pooling strategies \cite{desplanques_2020_ecapatdnn, zhu_2018_attnpool, okabe_2018_attnpool}, and margin-based training loss functions \cite{wang_2018_amsoftmax, deng_2019_aamsoftmax} for more discriminative speaker embeddings.

Meanwhile, the rise of self-supervised learning (SSL) transformers \cite{baevski_2020_wav2vec2, hsu_2021_hubert, chen_2022_wavlm} has spurred recent research into using pre-trained representations for SV.
Two main approaches have emerged: fine-tuning a pre-trained model into an end-to-end SV system \cite{fan_2021_explorewv2, vaessen_2022_finetunewv2}, or using SSL models as feature extractors for a downstream model \cite{chen_2022_wavlm, yang_2021_superb, chen_2022_largescale_for_asv, novoselov_2023_wav2vec2tdnn}.
Notably, the SUPERB benchmark \cite{yang_2021_superb} introduced a weighted sum of hidden states from all layers to produce downstream features.
\cite{chen_2022_wavlm} and \cite{chen_2022_largescale_for_asv} has achieved state-of-the-art SV performance by combining the SUPERB with a powerful ECAPA-TDNN \cite{desplanques_2020_ecapatdnn}.

However, current methods for exploiting pre-trained models in SV have limitations.
First, some still rely on the final layer's output \cite{fan_2021_explorewv2, vaessen_2022_finetunewv2, novoselov_2023_wav2vec2tdnn}, while several recent analyses \cite{chen_2022_wavlm, chen_2022_largescale_for_asv} have shown that speaker cues are concentrated in lower layers.
Second, most studies adopt the trivial layer aggregation from SUPERB \cite{yang_2021_superb}.
This static summation cannot capture frame-level variability of inter-layer importance, and its scalar weights are fixed after training, which limits generalization.

\begin{figure*}[ht]
\centering
\includegraphics[scale=0.5]{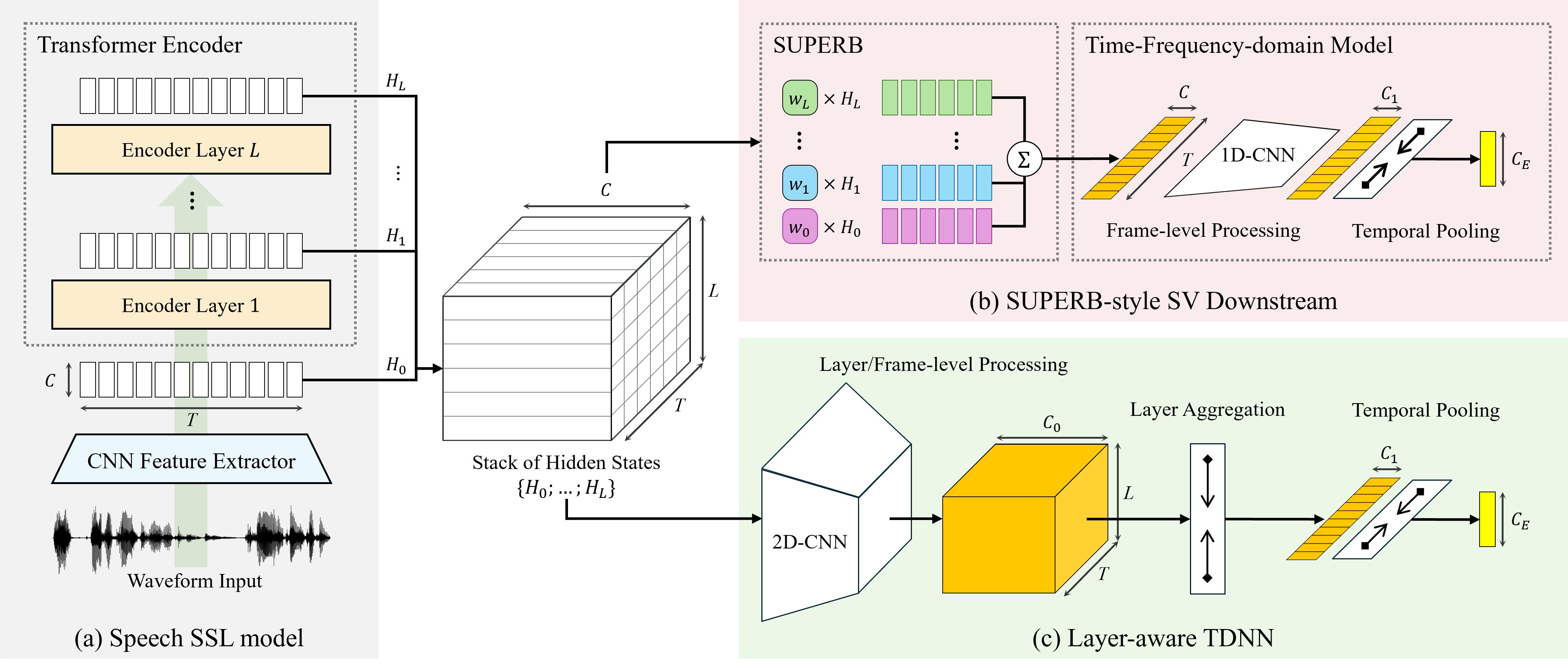}
\caption{Comparison of the speaker verification pipeline leveraging the multi-layer features from the speech SSL model.}
\label{fig1:overview_pipeline}
\end{figure*}

To overcome these limitations, we propose a dedicated backend architecture that fully exploits multi-layer SSL representations.
Our contributions are summarized as three-fold:
\textbf{(1)} we introduce a layer-aware TDNN (L-TDNN) that operates directly on stacked hidden states to enrich speaker characteristics;
\textbf{(2)} we devise a frame-adaptive attention pooling strategy for dynamic layer aggregation;
and \textbf{(3)} we validated comprehensive experiments that the proposed strategy consistently and efficiently outperforms existing methods for leveraging speech pre-trained models for SV.

\section{Related Work} \label{sec:related work}

\subsection{Leveraging Pre-trained Models for Speaker Verification}
\label{sec2:subsec:leveraging pre-trained models}
Starting from Wav2vec \cite{schneider_2019_wav2vec}, SSL Transformers \cite{baevski_2020_wav2vec2, hsu_2021_hubert, chen_2022_wavlm, baevski_2020_vq_wav2vec} have become widely accepted in contemporary speech-processing research.
These models aim to extract task-agnostic representations by directly processing raw waveforms and generally comprise convolution layers followed by a Transformer encoder.
Efforts to leverage such pre-trained models for SV have followed two main routes.

The first approach builds an end-to-end verification system by fine-tuning the SSL encoder.
Both \cite{fan_2021_explorewv2} and \cite{vaessen_2022_finetunewv2} explored adapting Wav2vec 2.0 \cite{baevski_2020_wav2vec2} for SV. 
\cite{fan_2021_explorewv2} formed an utterance-level representation by simply averaging the Transformer outputs and trained the network jointly on language and speaker classification.
\cite{vaessen_2022_finetunewv2} compared several pooling strategies for aggregating speaker information from the output sequence and proposed the insertion of a constant class (\textit{CLS}) token, which is inspired by BERT \cite{devlin_2019_bert}.

The other approach treats the SSL encoder as a frontend feature extractor so that an SV downstream model processes its output.
For example, \cite{novoselov_2023_wav2vec2tdnn} proposed a backend model, comprising two TDNN layers, statistic pooling, and a maxout linear layer, on top of Wav2vec 2.0 \cite{baevski_2020_wav2vec2}.
Although the proposal of SUPERB \cite{yang_2021_superb} was not exclusive to SV, its idea of combining hidden states from each layer of the pre-trained model inspired many subsequent studies.
While SUPERB adopted the $x$-vector \cite{snyder_2018_xvector} to process a weighted summation of layer-wise representations, \cite{chen_2022_wavlm} and \cite{chen_2022_largescale_for_asv} considered a more powerful off-the-shelf downstream architecture, ECAPA-TDNN \cite{desplanques_2020_ecapatdnn}, to achieve strong verification performances.

\subsection{Limitations of Prior Works and Preliminaries}
As surveyed above, recent SV studies have shifted to the SSL paradigm, guaranteeing faster convergence and strong downstream performance.
A diverse array of strategies has been discussed to exploit the pre-trained encoders, yet these approaches still face notable limitations.

While end-to-end approaches represent the speaker solely with the final layer output, the layer-wise probing across diverse SSL models \cite{chen_2022_wavlm, chen_2022_largescale_for_asv, ashihara_2024_sslanalysis, chen_2022_sslanalysis} demonstrated that speaker cues are prominent at lower layers.
Adapting the pre-trained encoder in an end-to-end manner, therefore, starts at a disadvantage for speaker discrimination.
SUPERB-based systems mitigate this issue by incorporating multi-layer hidden states.
However, its static and global-constant aggregation ignores frame-level variability and limits the capability of layer-wise representation.

These findings lead to motivation for the next step for SV in leveraging pre-trained models, a backend that can actively process the entire stack of layer-wise hidden states.
Conventional downstream networks, designed for the time-frequency domain, are not equipped to handle this additional layer dimension. 
Therefore, in this study, we introduce a dedicated backend for SSL encoders, which refines speaker features across layers and frames, and also dynamic aggregation of each dimension.

\section{Speaker Extraction using SSL Transformers} \label{sec:methodology}

Fig. \ref{fig1:overview_pipeline} provides an overview of the processing pipelines discussed in this section.
(a) draws the common architecture of contemporary speech SSL models such as Wav2vec 2.0 \cite{baevski_2020_wav2vec2}, HuBERT \cite{hsu_2021_hubert}, and WavLM \cite{chen_2022_wavlm}.
A convolutional (CNN) feature extractor first produces a latent representation $H_{0} \in \mathbb{R}^{C \times T}$ directly from the raw waveform, then Transformer layers $\{1, ..., L\}$ process the feature.
The hidden states from each layer are stacked, forming the initial input tensor $\{H_{0}; ...; H_{L}\} \in \mathbb{R}^{C \times T \times L}$ for the SV downstream.

Afterwards, parts (b) and (c) demonstrate how the SUPERB-based downstream and the proposed L-TDNN deal with the given tensor, respectively.
As (b) illustrates, earlier studies opted for the conventional time-frequency-based backends to process the SSL model features; therefore, they adopted the static weighted summation strategy to integrate the layer dimension in advance.
On the other hand, (c) depicts the proposed L-TDNN transforming the tensor into a speaker embedding, where we design the network to process the input tensor directly.
L-TDNN comprises three stages: a convolutional processing network at the layer and frame level, a layer aggregation layer, and a temporal pooling layer.

\subsection{Layer and Frame-level Processing Network}

\begin{figure}[!t]
\centering
\includegraphics[scale=0.55]{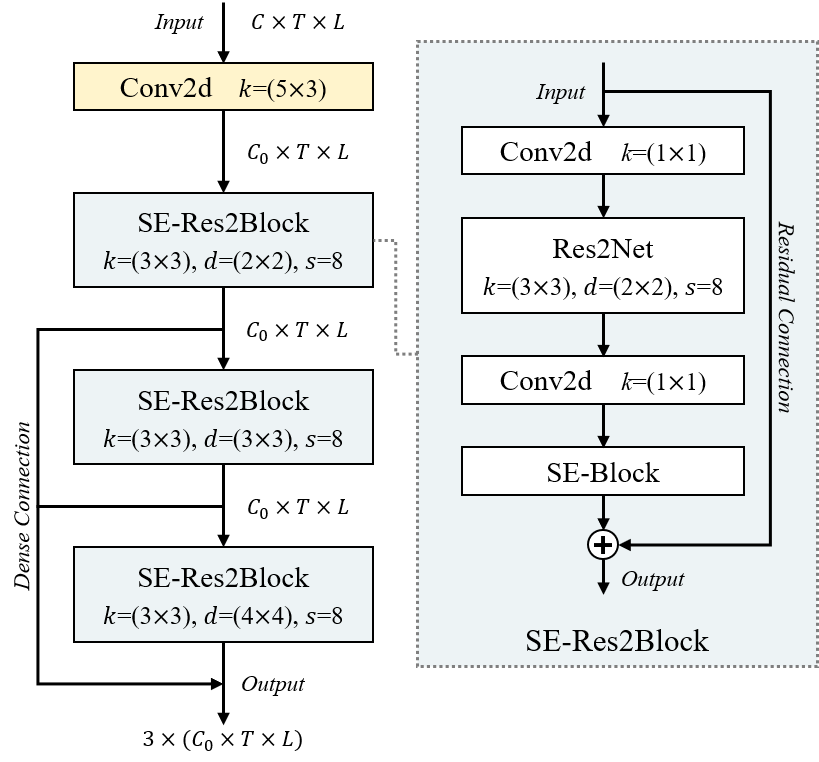}
\caption{Details of the layer/frame-level processing network.}
\label{fig2:layer_frame_level_processing_network}
\end{figure}

We extend the architecture from one of the powerful downstream models, ECAPA-TDNN \cite{desplanques_2020_ecapatdnn}, so that the network operates on a two-dimensional feature map.
SE-Res2Block, proposed from ECAPA-TDNN, benefits from combining the Squeeze-and-Excitation (SE) \cite{hu_2018_seblock} and the Res2Net \cite{gao_2019_res2net} module that processes multi-scale features through the hierarchical residual connections.
Moreover, the dense connection \cite{huang_2017_densenet} of each SE-Res2Block enables the shallow layers to contribute to a stronger foundation of speaker embedding.
Fig. \ref{fig2:layer_frame_level_processing_network} depicts the details of the convolutional network and its feature maps.
$C$, $T$, and $L$ denote the hidden size, the number of frames, and the number of SSL model hidden layers composing the input tensor $\{ H_{0}, ..., H_{L} \}$.
$k$, $d$, and $s$ are arguments for the convolutional operations, which are the kernel size, dilation, and scale at the Res2Net module, respectively.
During the expansion, we adjust the hidden size of the convolutional topology, $C_0=256$, to be smaller than the original's, keeping the number of parameters in L-TDNN comparable to SUPERB-style approaches \cite{chen_2022_wavlm, chen_2022_largescale_for_asv}.

\subsection{Frame-adaptive Layer Aggregation}
The following steps involve aggregating and pooling the feature map into a single vector representation of the speaker embedding.
To better exploit the rich information across multiple layers, we explore a more advanced strategy for aggregating layer-wise speaker cues than static weighted summation.
Inspired by one designed for multi-sequence aggregative processing \cite{kim_2023_way}, we devise a layer pooling layer based on SE \cite{hu_2018_seblock} combined with multi-head projection \cite{vaswani_2017_transformer}.

\begin{figure}[!t]
\centering
\includegraphics[scale=0.55]{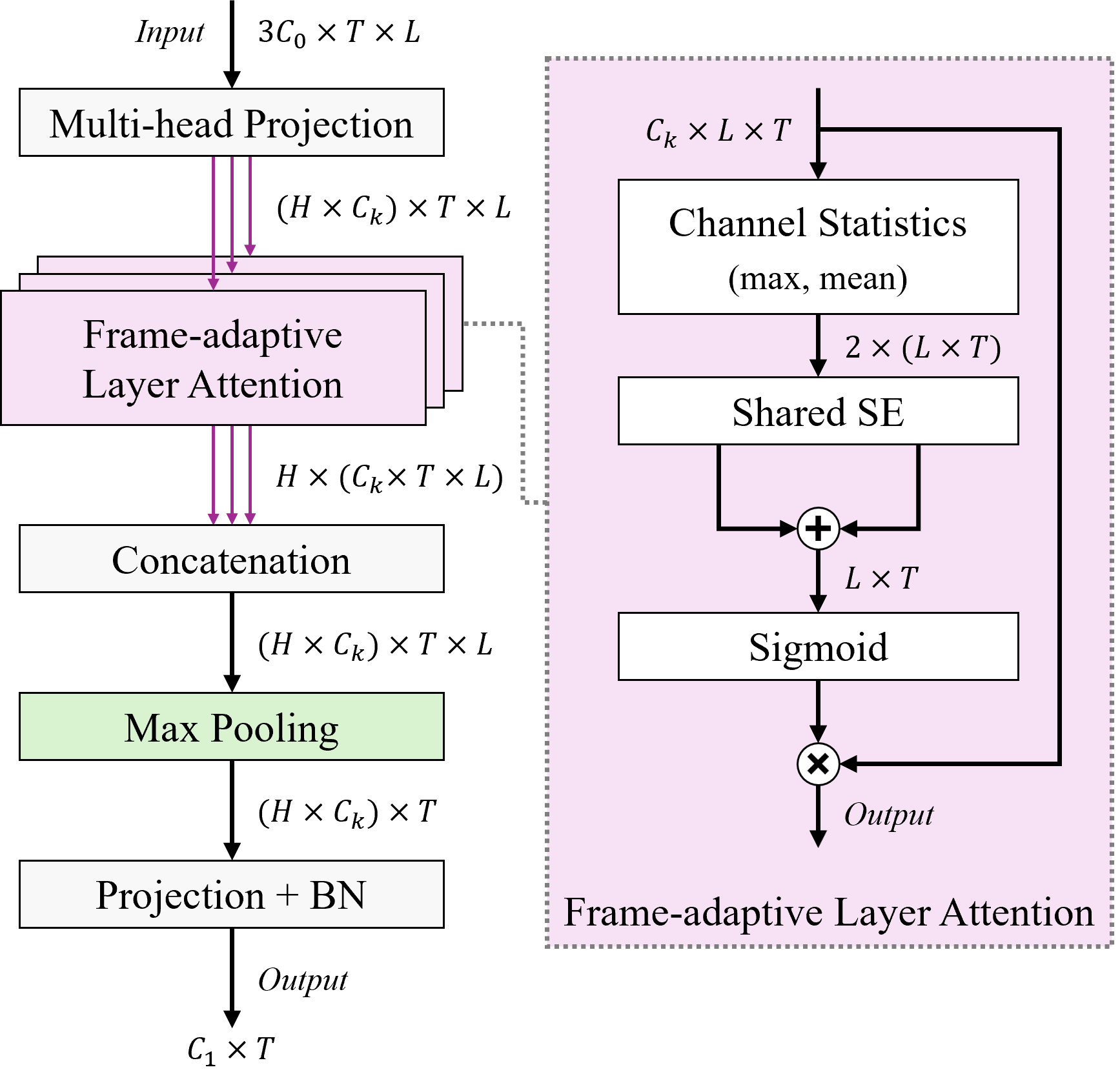}
\caption{Layer aggregation based on the MCA \cite{kim_2023_way} strategy.}
\label{fig3:layer_aggregation}
\end{figure}

Fig. \ref{fig3:layer_aggregation} illustrates the proposed layer-aggregation strategy, where the entire sequence of operations is shared across frames, and we achieve the frame-dependent usage of the layer dimension.
Given an output $X \in \mathbb{R}^{3C_0 \times T \times L}$ from the preceding convolution network, the pooling process comprises three steps.
First, we project the channel dimension with a learnable matrix $W_{in} \in \mathbb{R}^{C_k \times 3C_0}$:
\begin{equation}
    \begin{aligned}
    x &= W_{in} \cdot X
    \end{aligned}
\end{equation}
where $x \in \mathbb{R}^{C_k \times T \times L}$ represents one of the $H$ head projections.
In this study, we set $H=8$ and $C_k = \frac{3C_0}{H}.$

Then, we compute layer-wise weights through the SE operation for each head.
We start with taking the maximum and mean over the latent dimension, $x_{\text{max}}, x_{\text{mean}} \in \mathbb{R}^{L \times T}$. 
These statistics pass through the learnable parameters $W_{sq} \in \mathbb{R}^{r \times L}$ and $W_{sq} \in \mathbb{R}^{L \times r}$ as:
\begin{equation}
    \begin{aligned}
    \alpha_{z} &= W_{ex} \cdot \text{ReLU}(W_{sq} \cdot z)  \\
    \alpha &= \sigma( \sum_z \alpha_z), \hspace{10pt} z \in \{x_{\text{max}}, x_{\text{mean}}\} \\
    \end{aligned}
\end{equation}
where $r=\frac{1}{2}L$, $\sigma(\cdot)$ denotes the sigmoid activation that maps latent values within 0--1, and $\alpha \in \mathbb{R}^{L \times T}$ implies the layer importance at each frame. 

At last, we pool the most salient cues over layers by pooling the maximum after applying the weights.
\begin{equation}
    \begin{aligned}
    x &= \max_{L}(\alpha \odot x).
    \end{aligned}
\end{equation}
As above, we define the process within a single head, followed by head-wise concatenation and projection through the parameter $W_{out} \in \mathbb{R}^{C_1 \times (H \times C_k)}$ where $C_1=512$.

\subsection{Attentive Statistic Pooling}
The self-attention mechanism has proven to be successful in aggregating speaker embeddings from a sequence of frame-level features \cite{desplanques_2020_ecapatdnn, zhu_2018_attnpool, okabe_2018_attnpool}.
We adopt the strategy of ECAPA-TDNN \cite{desplanques_2020_ecapatdnn}, which uses the concept of the weighted mean and standard deviation for channel-dependent statistics.

Given the input $X \in \mathbb{R}^{C_1 \times T}$, the attention mechanism estimates $\alpha \in \mathbb{R}^{C_1 \times T}$, which implies the importance of each frame for its channel-wise statistics. 
Therefore, we compute the weighted statistics by:
\begin{equation}
    \begin{aligned}
    \hat{\mu} &= \sum^T_t \alpha_{c, t} \cdot X_{c, t} \\
    \hat{\sigma} &= \sqrt{\sum^T_t \alpha_{c, t} \cdot (X_{c, t})^2 - (\hat{\mu}_c)^2}.
    \end{aligned}
\end{equation}
We concatenate the attention-weighted statistics $\Tilde{\mu}, \Tilde{\sigma} \in \mathbb{R}^{C}$, and the linear transformation $W \in \mathbb{R}^{2C_1 \times C_E}$ and batch normalization follow, where $C_1=512$ and $C_E=192$.

\begin{table*}[!t]
    \caption{Evaluation with direct fine-tuning approaches in diverse training environments}
    \label{tab1:evaluation_in_diverse_environments}
    \centering
    \resizebox{0.93\textwidth}{!}{
        \begin{tabular}{l ccc c ccc c ccc c ccc}
            \toprule
            \multicolumn{1}{c}{\multirow{2.5}{*}{Verification System}} 
                & \multicolumn{3}{c}{VCTK} && \multicolumn{3}{c}{LibriSpeech} &
                    & \multicolumn{3}{c}{VoxCeleb1} && \multicolumn{3}{c}{VoxCeleb2} \\
                \cmidrule{2-4} \cmidrule{6-8} \cmidrule{10-12} \cmidrule{14-16}
                & EER* & EER & minDCF && EER* & EER & minDCF &
                    & EER* & EER & minDCF && EER* & EER & minDCF \\
            \midrule
            \multicolumn{4}{l}{Input: \textbf{MFCC}} \\
            \specialrule{0pt}{1pt}{2pt}
            \hspace{6pt} \textit{x}-vector 
                & 16.15 & 16.22 & 0.898 && 8.08 & 7.99 & 0.468 &
                & 12.07 & 12.08 & 0.742 && 7.94 & 7.97 & 0.490 \\
            \hspace{6pt} ECAPA-TDNN 
                & 5.44 & 5.35 & 0.585 && 2.45 & 1.91 & 0.117 &
                & 5.52 & 5.28 & 0.486 && 2.44 & 2.34 & 0.154 \\
            \midrule
            \midrule

    \multicolumn{16}{l}{\footnotesize\selectfont * \textsc{BASE} models, pre-trained on 960 hrs (LibriSpeech)} \\
    \specialrule{0pt}{1pt}{2pt}
            \textbf{Wav2vec 2.0} & \\
            \specialrule{0pt}{1pt}{2pt}
            \hspace{6pt} Temporal mean
                & 6.64 & 6.64 & \textbf{0.758} &
                & 3.03 & 2.54 & 0.174 &
                & 7.88 & 7.83 & 0.621 && 4.63 & 4.71 & 0.322 \\
            \hspace{6pt} [$CLS$] insertion
                & 9.77 & 9.46 & 0.980 && 3.40 & 2.79 & 0.175 &
                & 3.73 & 3.69 & 0.384 &
                & \textbf{2.25} & 2.27 & 0.187 \\
            \rowcolor{gray!20} 
            \hspace{6pt} \textbf{L-TDNN} (proposed)
                & \textbf{3.51} & \textbf{3.58} & 0.773 &
                & \textbf{1.83} & \textbf{1.16} & \textbf{0.095} &
                & \textbf{2.58} & \textbf{2.38} & \textbf{0.243} &
                & \textbf{2.25} & \textbf{2.19} & \textbf{0.144} \\
            \specialrule{0pt}{3pt}{3.5pt}
    
            \textbf{HuBERT} \\
            \specialrule{0pt}{1pt}{2pt}
            \hspace{6pt} Temporal mean
                & 5.56 & 5.44 & \textbf{0.591} &
                & 26.12 & 26.17 & 0.717 &
                & 32.81 & 33.73 & 0.888 && 32.72 & 31.17 & 0.713 \\
            \hspace{6pt} [$CLS$] insertion
                & 24.02 & 24.20 & 0.997 && 12.21 & 12.03 & 0.772 &
                & 22.84 & 22.65 & 0.991 && 15.59 & 15.48 & 0.933 \\
            \rowcolor{gray!20} 
            \hspace{6pt} \textbf{L-TDNN}
                & \textbf{3.78} & \textbf{3.84} & 0.599 &
                & \textbf{1.38} & \textbf{0.95} & \textbf{0.062} &
                & \textbf{2.42} & \textbf{2.23} & \textbf{0.211} & 
                & \textbf{2.00} & \textbf{1.95} & \textbf{0.135} \\
            \specialrule{0pt}{3pt}{3.5pt}

            \textbf{WavLM} \\
            \specialrule{0pt}{1pt}{2pt}
            \hspace{6pt} Temporal mean
                & 4.63 & 4.59 & \textbf{0.451} &
                & 27.29 & 27.20 & 0.773 &
                & 33.01 & 33.93 & 0.890 &
                & 30.11 & 29.53 & 0.798 \\
            \hspace{6pt} [$CLS$] insertion
                & 25.80 & 25.66 & 0.995 && 12.07 & 11.89 & 0.794 &
                & 20.99 & 20.85 & 0.989 && 9.89 & 9.81 & 0.728 \\
            \rowcolor{gray!20} 
            \hspace{6pt} \textbf{L-TDNN}
                & \textbf{3.43} & \textbf{3.55} & 0.555 &
                & \textbf{1.61} & \textbf{1.21} & \textbf{0.081} &
                & \textbf{2.15} & \textbf{1.96} & \textbf{0.218} & 
                & \textbf{1.94} & \textbf{1.88} & \textbf{0.121} \\
            \bottomrule
        \end{tabular}
    }
\end{table*}

\section{Experiment} \label{sec:experiment}

\subsection{Datasets and Implementation Details}
\label{section:experimental_details:subsection:datasets}

We evaluated our systems on multiple datasets to cover diverse training scenarios.
VCTK CSTR Corpus (VCTK) \cite{yamagishi19_vctk_corpus} provides clean recordings from 108 English speakers.
LibriSpeech \cite{panayotov15_librispeech} contains about $1\text{,}000$ hours of speech from $2\text{,}484$ speakers.
Finally, the VoxCeleb 1 \& 2 datasets \cite{nagrani17_voxceleb1, chung18b_voxceleb2} offer speech data within a variety of acoustic environments and noises by $1\text{,}369$ and $6\text{,}152$ speakers, respectively.
All audio was resampled to 16kHz.

Our model was trained with AAM-softmax \cite{deng_2019_aamsoftmax} with scale $30$ and margin $0.2$.
Adam optimizer \cite{kingma15_adamoptim} was adopted with a one-cycle learning rate ($lr$) scheduling \cite{smith19_onecyclelr} with the maximum $lr=0.003$ using an initial $10\%$ warmup phase.
Minibatches comprise $128$ samples, and each is truncated to $3$s.
We implemented data augmentation to drop a random span from the SSL model output, inspired by SpecAugment \cite{park19_specaugment} and Patchout \cite{koutini_2022_passt}.
On evaluation, we used cosine similarity.

Throughout our experiments, SSL models were kept frozen while they served as feature extractors, and no score calibration was applied.
This setup aims to isolate and validate the effectiveness of the proposed layer-aware processing on leveraging pre-trained features.
We further discuss the joint training of the SSL frontend and the speaker backend and also post-processing techniques in Section \ref{sec:conclusion}.

\subsection{Evaluation Metrics}
\label{section:experimental_details:subsection:evaluation_metrics}
We use two general SV evaluation metrics: equal error rate (EER) and minimum detection cost function (minDCF), with minDCF parameters set to $C_\text{Miss}=1$ and $C_\text{FA}=1$, and $P_\text{target}=0.01$.
Both work by finding optimal decision threshold based on the similarity scores of a given evaluation set.

For a more practical evaluation, we introduce an additional measure, EER*, where the test set is unseen during threshold setting.
As below, EER* evaluates the test set ($\mathcal{Z}$) using a fixed threshold ($\tau$) derived from the EER of the validation set ($\mathcal{U}$).
It takes the mean of the resulting test set FAR and FRR.
\begin{equation}
\label{evaluation_metric:equal_error_rate_star}
\begin{aligned}
\mathrm{EER}^{*} &= \frac{\mathrm{FAR}_{\mathcal{Z}}(\tau) + \mathrm{FRR}_{\mathcal{Z}}(\tau)}{2} \\
\text{s.t.\space\space} &\mathrm{EER}_{\mathcal{U}} = \mathrm{FAR}_{\mathcal{U}}(\tau) = \mathrm{FRR}_{\mathcal{U}}(\tau)
\end{aligned}
\end{equation}

\subsection{Baselines and Pre-trained SSL Models}
We established two baseline groups that use pre-trained models.
The first was to fine-tune an SSL model and pool a speaker embedding from the final hidden layer using non-parameterized ways.
Either temporal mean pooling \cite{fan_2021_explorewv2} or a CLS token insertion \cite{vaessen_2022_finetunewv2} was adopted.
The second group followed the SUPERB pipeline, which uses a weighted sum of multi-layer features to produce an input for the downstream SV model.
We paired this strategy with $x$-vector \cite{snyder_2018_xvector} and ECAPA-TDNN \cite{desplanques_2020_ecapatdnn}.
Experiments were conducted using three representative speech SSL models: Wav2vec 2.0 \cite{baevski_2020_wav2vec2}, HuBERT \cite{hsu_2021_hubert}, and WavLM \cite{chen_2022_wavlm}.

\subsection{Experimental Results}
\subsubsection{\textbf{Comparison with End-to-end Fine-tuning Approaches}}
Table \ref{tab1:evaluation_in_diverse_environments} compares L-TDNN against two end-to-end fine-tuning approaches as well as traditional MFCC-based verification systems.
Across all three SSL encoders and every evaluation corpus, L-TDNN consistently demonstrates superior performance, achieving the lowest EER and EER* in all conditions.

In contrast, the fine-tuning baselines show poor generalization, exhibiting instability with SSL Transformers other than Wav2vec 2.0, likely due to differing pre-training objectives.
Notably, the temporal mean pooling \cite{fan_2021_explorewv2} often performed worse than the MFCC-based ECAPA-TDNN, while the CLS insertion \cite{vaessen_2022_finetunewv2} has only proved highly limited model and dataset combinations.
These results reveal the stability of leveraging multi-layer features from pre-trained models, and vice versa, the high sensitivity of naive fine-tuning while relying on the final layer output.
L-TDNN's stable and superior performance supports its approach, benefiting from diverse types of speech SSL models.

\subsubsection{\textbf{Comparison with SUPERB-based Approaches}}

Table \ref{tab2:comparison_with_feature_extractor_approaches} compares L-TDNN against SUPERB-based baselines paired with $x$-vector \cite{snyder_2018_xvector} and ECAPA-TDNN \cite{desplanques_2020_ecapatdnn} backends, using various scales of SSL frontends.
L-TDNN surpasses the baselines across all evaluation metrics, regardless of the scale of a pre-trained encoder.
On average, L-TDNN achieves an EER* improvement of $45\%$ over the $x$-vector baseline and $14\%$ over the stronger ECAPA-TDNN baseline.

Furthermore, L-TDNN demonstrates strong scalability as the pre-training data and model capacity increase.
It maintains a significant performance margin over the ECAPA-TDNN with relative EER* improvements of $9\%$ (WavLM base+) and $15\%$ (WavLM large); $9\%$ and $11\%$ on EER, respectively.
It is encouraging that L-TDNN yields significant gains over the architecturally similar ECAPA-TDNN.
This highlights the effectiveness of modeling inter-layer information from hidden states of pre-trained models for the downstream task.

\begin{table}[!t]
    \caption{Comparison with SUPERB-based SV systems}
    \label{tab2:comparison_with_feature_extractor_approaches}
    \centering
    \resizebox{\linewidth}{!}{
        \begin{tabular}{l@{\quad} ccc c ccc}
            \toprule
    \multicolumn{1}{c}{\multirow{2.5}{*}{Verification System}} 
        & \multicolumn{3}{c}{VoxCeleb1} && \multicolumn{3}{c}{VoxCeleb2} \\
        \cmidrule{2-4} \cmidrule{6-8} 
        & EER* & EER & minDCF && EER* & EER & minDCF \\
            \midrule

    \multicolumn{8}{l}{\footnotesize\selectfont * \textsc{BASE} models, pre-trained on 960 hrs (LibriSpeech)} \\
    \specialrule{0pt}{1pt}{2pt}
            \textbf{Wav2vec 2.0} & \\
            \specialrule{0pt}{1pt}{2pt}
            \hspace{3pt} $x$-vector
                & 4.64 & 4.65 & 0.468 && 4.08 & 3.95 & 0.351 \\
            \hspace{3pt} ECAPA-TDNN
                & 3.24 & 2.82 & 0.346 && 2.42 & 2.33 & 0.157 \\
            \rowcolor{gray!20} 
            \hspace{3pt} \textbf{L-TDNN} (proposed)
                & \textbf{2.58} & \textbf{2.38} & \textbf{0.243} &
                & \textbf{2.25} & \textbf{2.19} & \textbf{0.144} \\
            \specialrule{0pt}{3pt}{3.5pt}

            \textbf{HuBERT} \\
            \specialrule{0pt}{1pt}{2pt}
            \hspace{3pt} $x$-vector
                & 4.00 & 3.97 & 0.423 && 3.66 & 3.43 & 0.305 \\
            \hspace{3pt} ECAPA-TDNN
                & 3.18 & 2.53 & 0.309 && 2.24 & 2.06 & 0.149 \\
            \rowcolor{gray!20} 
            \hspace{3pt} \textbf{L-TDNN}
                & \textbf{2.42} &\textbf{ 2.23} & \textbf{0.211} &
                & \textbf{2.00} & \textbf{1.95} & \textbf{0.135} \\
            \specialrule{0pt}{3pt}{3.5pt}

            \textbf{WavLM} \\
            \specialrule{0pt}{1pt}{2pt}
            \hspace{3pt} $x$-vector
                & 4.01 & 3.97 & 0.428 && 3.73 & 3.61 & 0.325 \\
            \hspace{3pt} ECAPA-TDNN
                & 2.55 & 2.15 & 0.257 && 2.07 & 2.05 & 0.137 \\
            \rowcolor{gray!20} 
            \hspace{3pt} \textbf{L-TDNN}
                & \textbf{2.15} & \textbf{1.96} & \textbf{0.218} & 
                & \textbf{1.94} & \textbf{1.88} & \textbf{0.121} \\
            \midrule

    \multicolumn{8}{l}{\footnotesize\selectfont * \textsc{BASE} model, pre-trained on 94K hrs$^\dagger$} \\
    \specialrule{0pt}{1pt}{2pt}
            \textbf{WavLM+} & \\
            \specialrule{0pt}{1pt}{2pt}
            \hspace{3pt} ECAPA-TDNN
                & 2.21 & 1.84 & 0.217 &
                & 1.79 & 1.72 & 0.094 \\
            \rowcolor{gray!20} 
            \hspace{3pt} \textbf{L-TDNN}
                & \textbf{1.93} & \textbf{1.62} & \textbf{0.198} &
                & \textbf{1.70} & \textbf{1.63} & \textbf{0.092} \\
            \midrule

    \multicolumn{8}{l}{\footnotesize\selectfont * \textsc{LARGE} model, pre-trained on 94K hrs$^\dagger$} \\
    \specialrule{0pt}{1pt}{2pt}
            \textbf{WavLM} &  \\
            \specialrule{0pt}{1pt}{2pt}
            \hspace{3pt} ECAPA-TDNN
                & 2.19 & 1.78 & 0.213 && 1.79 & 1.72 & 0.110 \\
            \rowcolor{gray!20} 
            \hspace{3pt} \textbf{L-TDNN}
                & \textbf{1.85} & \textbf{1.51} & \textbf{0.167} &
                & \textbf{1.54} & \textbf{1.61} & \textbf{0.104} \\
            \bottomrule
            \specialrule{0pt}{0pt}{1pt}
            \multicolumn{8}{l}{\small\selectfont $^{\dagger}$Libri-Light, GigaSpeech, and VoxPopuli} \\
        \end{tabular}
    }
\end{table}

\subsubsection{\textbf{Analyses on Model Efficiency}}
\begin{figure}[!t]
\centering
\includegraphics[scale=0.55]{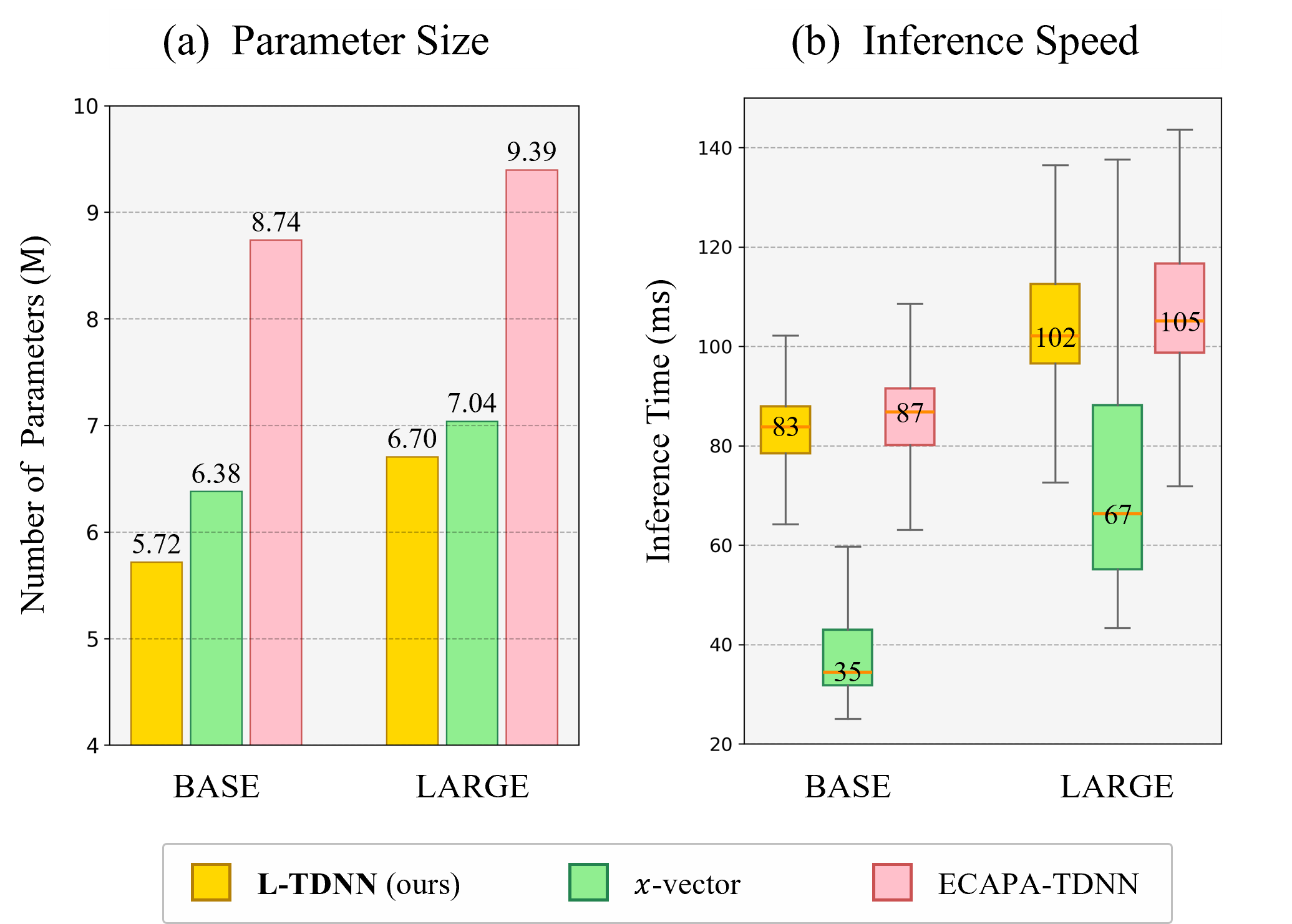}
\caption{Comparison of model efficiency across speaker embedding backends and SSL-frontend scales.}
\label{fig4:efficiency_comparison}
\end{figure}

Fig. \ref{fig4:efficiency_comparison} analyzes the efficiency of L-TDNN against SUPERB-based systems by comparing (a) parameter counts and (b) inference latency.
We measured the latency on ``Vox-O'' trials using a single NVIDIA RTX A6000 GPU, reporting the median.
L-TDNN is the most parameter-efficient model, being approximately two-thirds the size of ECAPA-TDNN while delivering the better performance shown in Table \ref{tab2:comparison_with_feature_extractor_approaches}.
Regarding inference speed, L-TDNN also holds a slight advantage over ECAPA-TDNN.
Although the simpler $x$-vector backend is faster, it suffers a significant trade-off in verification performance.
These results demonstrate that L-TDNN offers a competitive advantage of accuracy, compactness and efficiency in leveraging features from the pre-trained Transformers.

\section{Conclusion} \label{sec:conclusion}

\subsection{Limitations and Future Works}
In this study, we froze the SSL model parameters to isolate our backend's contribution.
A key direction for future work is to explore joint training of the SSL frontend and the speaker backend, unleashing more powerful, task-specific representations.
Additionally, we will investigate the integration of common post-processing techniques, such as score calibration \cite{chen_2022_wavlm, qmf_2021_lmft}, to further push the model's verification performance toward the state-of-the-art in SV benchmarks.

\subsection{Conclusion}
This study introduced L-TDNN, a novel backend architecture designed to effectively leverage the multi-layered nature of pre-trained speech encoders for speaker verification.
L-TDNN directly addresses the layer dimension, given the stack of hidden states produced from each layer of an SSL Transformer, which is previously overlooked.
It is achieved through a dedicated structure comprising a layer-aware convolutional network, a frame-adaptive layer aggregation, and attentive temporal pooling, allowing it to robustly model inter-layer speaker characteristics.

Through extensive experiments, we demonstrated that L-TDNN consistently outperforms primary existing approaches such as end-to-end fine-tuning and SUPERB-based feature extraction.
This strong performance was validated across diverse SSL models (Wav2vec 2.0, HuBERT, VoxCeleb 1 \& 2).
Moreover, we verified that L-TDNN provides these performance gains while also being more parameter-efficient and computationally faster than comparable backend architectures.
These findings confirm L-TDNN as a robust, generalizable, and efficient solution, highlighting the benefits of dedicated layer-aware processing for speaker verification.
Future work will incorporate jointly training of the SSL frontend and the speaker backend, as well as score calibration techniques, to further push the system beyond its current capabilities.

\newpage
\bibliography{reference}
\bibliographystyle{IEEEtran}

\end{document}